\documentclass[traditabstract]{aa}
\usepackage{graphicx}
\usepackage{txfonts}
\usepackage{natbib}
\usepackage{enumitem}
\bibpunct{(}{)}{;}{a}{}{,}

 \newcommand{\mic}{$\mu$m}
 \newcommand{\mics}{$\mu$m~}

\begin{document}

\title{The evolving slope of the stellar mass function at $0.6 \leq z < 4.5$ from deep WFC3 data} 

\author{
P. Santini\inst{1}
\and
A. Fontana\inst{1}
\and
A. Grazian\inst{1}
\and
S. Salimbeni\inst{2}
\and
F. Fontanot\inst{3}
\and
D. Paris\inst{1}
\and
K. Boutsia\inst{1}
\and
M. Castellano\inst{1}
\and
F. Fiore\inst{1}
\and
S. Gallozzi\inst{1}
\and
E. Giallongo\inst{1}
\and
A. M. Koekemoer\inst{4}
\and
N. Menci\inst{1}
\and
L. Pentericci\inst{1}
\and
R. S. Somerville\inst{4,5}
}

  \offprints{P. Santini, \email{paola.santini@oa-roma.inaf.it}}

\institute{INAF - Osservatorio Astronomico di Roma, via di Frascati 33, 00040 Monte Porzio Catone, Italy.
\and Astronomy Department, University of Massachusetts, Amherst, MA 01003, U.S.A.
\and INAF - Osservatorio Astronomico di Trieste, Via G.B. Tiepolo 11, 34131 Trieste, Italy.
\and Space Telescope Science Institute, 3700 San Martin Drive, Baltimore, MD 21218.
\and Department of Physics and Astronomy, Johns Hopkins University, Baltimore, MD 21218.
}

   \date{Received .... ; accepted ....}
   \titlerunning{The evolving slope of the stellar mass function at $0.6 \leq z < 4.5$ from deep WFC3 observations}

   \abstract{We used Early Release Science (ERS) observations taken with the Wide Field Camera 3 (WFC3) in the  GOODS-S field to study the galaxy stellar mass function (GSMF) at $0.6 \leq z < 4.5$. Deep WFC3 near-IR data (for $Y$ as faint as 27.3, $J$ and $H$ as faint as 27.4 AB mag at $5 \sigma$), as well as deep $K_S$ (as faint as 25.5 at $5 \sigma$) Hawk-I band data, provide an exquisite data set with which determine in an unprecedented way the low-mass end of the GSMF, allowing an accurate probe of masses as low as  $M_*\simeq 7.6 \cdot 10^9 M_\odot$ at $z\sim 3$.      
  Although the area used is relatively small ($\sim 33$ arcmin$^2$),  we   found generally good agreement with previous studies on the entire mass range. 
   Our results show that the slope of the faint-end  increases with redshift, from $\alpha = -1.44 \pm 0.03$ at $z\sim 0.8$ to $\alpha = -1.86 \pm 0.16$ at $z\sim 3$, although indications exist that it does not steepen further between $z\sim 3$ and $z\sim 4$. This result is insensitive to any uncertainty in the $M^*$ parameter.  
     The steepness of the GSMF faint-end  solves the well-known disagreement between the stellar mass density (SMD) and the integrated star formation history at $z>2$. However,     we confirm the that there appears to be an excess of integrated star formation  with respect to the SMD  at $z<2$, by a factor of $\sim 2-3$. 
Our comparison of the observations with theoretical predictions shows that the models forecast a  greater abundance of low mass galaxies, at least up to $z\sim 3$, as well as a dearth of massive galaxies at $z\sim 4$ with respect to the data, and that the predicted SMD is generally overestimated at $z\lesssim 2$.
}


\maketitle

\section{Introduction}\label{sec:intro}

Understanding the assembly of stellar mass in galaxies 
is a fundamental step towards a description of  galaxy evolution. 
Key tools to study this process through cosmic time are the galaxy stellar mass function (GSMF) and its integral over masses (the stellar mass density, SMD hereafter). 

Therefore, it is no surprise that most extragalactic surveys in the past decade have been used to determine the shape and evolution of the GSMF as a function of redshift. 
The earliest results based on small field surveys revealed that the SMD decreases with increasing redshift, as expected in the framework of the currently accepted cosmological hierarchical scenario, both in terms of integrated SMD \citep[e.g.][]{giallongo98,dickinson03,fontana03,rudnick03}, as well as of the detailed GSMF \citep[e.g.][]{fontana04,drory04,drory05}.  
There have since been many indications  that the  evolution of the GSMF occurs more rapidly for more massive galaxies than for low mass ones \citep[e.g.][]{fontana06,pozzetti07,perezgonzalez08,kajisawa09,marchesini09}, a behaviour known as downsizing in stellar mass \citep[see][and references therein]{fontanot09}. 
The advent of near- and mid-infrared facilities, above all the  Spitzer telescope,  has allowed  the uncertainties in  stellar mass estimates to be reduced and the extension of their analysis  to $z\gtrsim3$ \citep[e.g.][]{fontana06,kajisawa09,caputi11}. 
In parallel, wide-field surveys have provided large samples with more accurate statistics \citep{drory09,pozzetti10,bolzonella10,marchesini10,ilbert10}. 
One of the key results of these surveys has been the demonstration that the shape of the GSMF  cannot be described by a (widely adopted) single Schechter function at least up to $z\simeq 1.5$, 
but that it departs from this parametric form  because of the superposition of individual distributions for the red and blue galaxy populations \citep{ilbert10,pozzetti10,bolzonella10,mortlock11} or/and because of a change with stellar mass either in star formation efficiency or galaxy assembly rate \citep{drory09}. 

An accurate knowledge of the GSMF is also a sensitive test of modern galaxy evolutionary models. From its initial studies, much interest in the GSMF has been triggered by the possibility of constraining the physics of the evolution of more massive galaxies, which, according to the hierarchical structure formation scenario, are the results of the merging of lower mass objects at earlier times \citep{cole94}. 
In addition,  to achieve a complete view of the galaxy formation picture, an important goal  is a robust knowledge of the properties of low mass galaxies at high redshift. 
The slope of the GSMF at low masses may also represent a critical benchmark for current galaxy formation models.  There is growing evidence that the number of low mass galaxies in the Universe is systematically overpredicted by most or all theoretical models \citep{fontana06,fontanot09,marchesini09}. Very similar evidence also appears  in the analysis of the luminosity functions \citep{poli01,marchesini07,lofaro09,henriques11}.  A particularly striking aspect of the mismatch is that it appears in  different renditions of theoretical models, suggesting that it marks some fundamental incompleteness in our theoretical understanding of galaxy formation and growth.

While a global picture is emerging from these investigations, many outstanding questions are still to be addressed.  
In general, the various GSMFs presented in the literature agree reasonably well at $z=0-5$, although disagreements exist, somewhat increasing at high redshift  \citep{caputi11,gonzalez11,marchesini10,mortlock11},  that cannot be explained  by merely field-to-field variance. At even higher redshift, the available estimates of the SMD are based on UV-selected samples, hence are potentially incomplete in mass, and/or are often derived by adopting average mass-to-light  ratios for the whole population rather than  detailed object-by-object estimates \citep{gonzalez11}. Finally, and particularly relevant for the main topic of this paper, the GSMF at low masses is highly uncertain at intermediate and high redshifts, since current samples do not extend to the depths required to establish its slope with good accuracy. 
These uncertainties are due to a number of observational limitations.

In addition to the uncertainties related to the GSMF computation, it must not be forgotten that the actual estimates of stellar masses from broad band photometry are potentially affected by many systematic uncertainties, even when accurate redshifts are available. Part of this uncertainty is due to the lack of knowledge of important parameters of the stellar population, such as metallicity or extinction curve. 
The modelling of highly uncertain phases of stellar evolution is another source of uncertainty: in particular the different treatments of the thermally pulsating asymptotic giant branch (TP-AGB) phase is the source of the highest discrepancies in simple stellar population models \citep[see e.g.][]{maraston05,marigo08}, and  has relevant implications for the estimate of near-infrared luminosities and stellar masses for galaxies dominated by intermediate-age stellar populations ($\sim 1$ Gyr). 
The largest bias is due to the difficulties in reconstructing the star formation history of each galaxy, which is necessary to estimate the appropriate $M_*/L$ ratio, and that may be poorly described by simplistic models such as those adopted in  stellar population synthesis codes \citep{maraston10,lee10}. 

All these uncertainties contribute to one of the main puzzles that appear in present-day observational cosmology: the mismatch between the observed SMD and the integrated star formation rate density (SFRD) \citep[e.g.][]{hopkins06,fardal07,wilkins08}. In principle, these two observables represent  independent approaches to studying the mass assembly history from different points of view. However, the integrated star formation, after considering the gas recycle fraction into the interstellar medium, appears to be higher than the observed SMD at all redshifts.  Several authors  have highlighted this severe discrepancy \citep[of up to a factor of $\sim 4$ at $z\sim 3$,][]{wilkins08}.  
Moreover, if the merging contribution to the stellar mass build-up is accounted for \citep{drory08}, the agreement gets even worse. 
Intriguingly, the integrated SFRD exceeds  the observed SMD, implying that we either overestimate the SFRD, or miss a substantial fraction of massive galaxies, or underestimate their masses, or finally fail in reconstructing the low-mass tail of the GSMF. An initial mass function (IMF) that varies over cosmic time was invoked to reconcile the two observables \citep{fardal07,wilkins08}. However, before invoking the non-universality of the IMF, it must be noted that both the SFRD and the SMD are affected by large uncertainties. 
The measure of the star formation rate is itself particularly difficult, being either highly dependent on uncertain dust corrections \citep[e.g.][]{santini09,nordon10} or limited to the brightest far-infrared galaxies at $z<2-3$ \citep{rodighiero10}. 
On the SMD side, in addition to the uncertainties related to the stellar mass estimate itself, a major role is played by the poor knowledge of the low-mass tail of the GSMF. 
Owing to the limited depths of current IR surveys, the estimate of the faint-end slope basically relies on large extrapolations. An incorrect estimate, given the large number density of low mass objects, could translate into non-negligible errors in the SMD.

A robust estimate of the slope of the GSMF is necessary to provide tighter constraints on all these unknowns. In this study we take advantage of the recent deep near-IR observations carried out by Wide Field Camera 3 (WFC3) installed on the HST in the upper part of the GOODS-S field in the $Y$, $J$ and $H$ bands and by Hawk-I mounted at VLT in the $K_S$ band.  These data allow accurate measurements of the stellar mass to very low limits.  In this respect, we extend to higher redshifts and lower masses the deep analysis carried out by \cite{kajisawa09}. The only study of comparable depth is  \cite{mortlock11}, which was also based on WFC3 data. 
However, the greater depth of the Early Release Science (ERS) images used in this work and the conservative cuts that we apply to the sample ensure an excellent overall photometric quality, as we discuss in Sect. \ref{sec:comparison}. 
Unfortunately, the area covered by ERS observations is  small compared to recent surveys, and is slightly overdense. 
This feature somewhat limits the universal validity of our results regarding the SMD, especially in the intermediate redshift bins, although we  chose our redshift intervals in order to ensure that the known clusters and groups (discussed in Sect. \ref{sec:data}) were mostly confined to two of them.  However, we show that the study of the faint-end slope, which is the main aim of the present analysis, is insignificantly  affected by these cosmic variance effects.   In addition, this work represents an  exercise to explore the potential of  future deep WFC3 observations, such as those  of the CANDELS survey \citep{grogin11,koekemoer11}, which will cover a much more extended area over various fields with depths comparable to the ERS observations.

The paper is organized as follows: 
after introducing the data in Sect. \ref{sec:data}, we  present the stellar mass estimate and the GSMF in  Sect. \ref{sec:mf}, the analysis of the faint-end slope in  Sect. \ref{sec:faintend}, the SMD and its comparison with the integrated SFRD in Sect. \ref{sec:md}, and the comparison with theoretical predictions in Sect. \ref{sec:models}. Sect. \ref{sec:summ} summarizes our results. 
In the following, we adopt the $\Lambda$-CDM concordance cosmological model (H$_0$ = 70 km/s/Mpc, $\Omega_{\small M} $ $=$\begin{scriptsize}\end{scriptsize} 0.3 and $\Omega_{\Lambda}$ = 0.7). All magnitudes are in the AB system.

\section{The data sample} \label{sec:data}

This work exploits a new set of near-IR images that represent a significant improvement in photometric quality and depth over existing surveys.  The first component is the public release of the Early Release Science (ERS) observations taken with WFC3, the new near-IR camera on board HST. The ERS observations cover an area of $\sim$ 50 arcmin$^2$, located in the northern $\sim$ 30\% of the GOODS-South field.  They were taken in three filters, $Y_{098}$, $J_{125}$, and $H_{160}$, which reach 27.3 ($Y$) and 27.4 ($J$, $H$) magnitudes at $5 \sigma$ in an area of $\sim$ 0.11 arcsec$^2$. We used the ERS mosaics produced as described in \cite{koekemoer11}; we also refer to \cite{grazian11} for details of the catalogues, and \cite{windhorst11} for a full description of the ERS observational program.

We  complemented these images with new deep $K_S$ band images taken over the GOODS-S field with the near-IR VLT imager Hawk-I. The latter were taken in the framework of a program designed to search for $z\sim 7$ galaxies \citep{castellano10a,castellano10b}. In the $K_S$ band, the surveyed area covers 80\% of the WFC3 ERS area. Owing to this and after excluding  image edges of dubious quality, the available area reduces to $\sim$ 33 arcmin$^2$. 
The data reduction of the $K_S$ images is analogous to the procedure used for other Hawk-I data \citep{castellano10a}.  The net exposure time is 25200s, with a $1 \sigma$ r.m.s. of 1.26 counts per second  in a 1'' aperture. The magnitude limit at $5 \sigma$ is $\sim25.5$, one magnitude deeper than the previous ISAAC $K_S$ band.

We  finally built a multiwavelength GOODS-ERS catalogue adding the other public images available in the GOODS-S field.  They include the ACS images in the $BVIz$ bands \citep{giavalisco04}, the deep $UR$ images from VIMOS \citep{nonino09} and the four IRAC bands at 3.6, 4.5, 5.8, and 8.0 \mic. With respect to the data set used to assemble our previous GOODS-MUSIC sample \citep{grazian06,santini09}, the present GOODS-ERS data set benefits not only from the much deeper IR coverage provided by the new WFC3 and $K_S$ band data, but also from a deeper version of the $z$ band image, which nearly doubles the exposure time of the previous image,  a deeper $U$ band image, and  a brand new $R$ image. 
In this data set, we  extracted a 14 band multiwavelength catalogue using the $H$ band as a  detection image. Colours were carefully obtained with the same technique used in the GOODS-MUSIC catalogue, where we adopted the PSF-matching code CONVPHOT \citep{desantis07} to accurately deblend objects in the ground-based and Spitzer images. We note that the depth of the $H$ band even exceeds the depth of the bluest bands, resulting in very poor quality photometric information about the faintest $H$-selected objects.

The catalogue was cross-correlated with existing spectroscopic samples. 
For sources lacking spectroscopic information, photometric redshifts were computed by fitting the 14 band multiwavelength photometry to the PEGASE 2.0 templates (\citealt{fioc97}, see details in \citealt{grazian06}). 
The accuracy reached by the photometric redshifts is very high, the absolute scatter $|\Delta z|/(1+z_{spec})$ being equal to 0.03, with only 3\% of severe outliers ($|\Delta z|/(1+z_{spec}) > 0.5$). 
The statistical error associated with each photometric redshift was used to evaluate the limiting magnitude at which a reliable GSMF can be computed. We found that a limit $H\simeq 26$ (or, equivalently, $K_S\simeq 25.5$) is appropriate to maintain the error in the mass estimate (see next section) to within 0.3 dex 
and the relative scatter in the photometric redshifts $|\Delta z|/(1 + z) < 0.1$ for 85\% of objects, and we adopt this in the following. 
From the analysis of the individual photometric-redshift probability distributions, we can compute the fraction of "reliable" candidates. We considered a candidate to safely lie within a given redshift interval when the integral of its probability distribution curve, normalized to unity, over that interval is larger than 90\%. Moreover, 
we accepted a certain level of tolerance in the definition of the redshift range to allow for the uncertainty in photometric redshifts. Following this method, for all $K_S<25.5$ sources with $z_{phot}>2$, we can exclude a secondary redshift solution at $z_{phot}<1.5$  in 97.2\% of the sources. This fraction increases to 99.6\%  when only bright sources ($K_s<24$) are considered.  
We also extracted a $K_S$ band detection catalogue, and verified that all the objects detected in the $K_S$ band are also detected in the $H$ one, which is unsurprising given the extraordinary quality of the WFC3 data. 

On the basis of these results, we decided to restrict our analysis to the $K_S\leq 25.5$ sample, albeit obtained from the $H$-selected one, for two reasons: firstly,  this selection allows a more robust comparison with previous $K$-selected surveys; and secondly, a $K_S = 25.5$ threshold is more efficient in detecting low mass objects than a $H =26$ one. Adopting this cut, we extend by two magnitudes the previous work of \cite{fontana06}, who studied the GSMF of the $K_S<23.5$ GOODS-MUSIC sample. 
Our $K_S\leq 25.5$ sample here includes 3210 objects, 421 of which have spectroscopic redshifts.

\begin{figure}[!t]
\resizebox{\hsize}{!}{\includegraphics[angle=270]{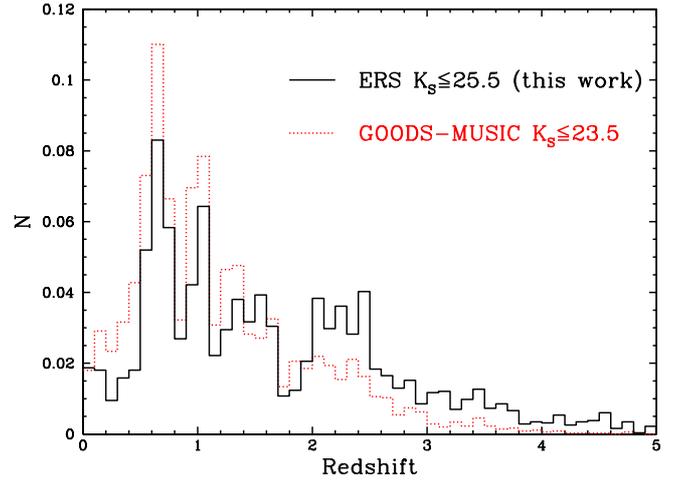}}
\caption{Redshift distribution of the $K_S\leq 25.5$  GOODS-ERS sample (black solid histogram) compared to the $K_S\leq 23.5$ GOODS-MUSIC one adopted by \cite{fontana06} (red dotted histogram). Overdensities at $z\simeq 0.7$, $z\simeq 1$, $z\simeq 1.6$, and $z\simeq 2.2-2.3$ can be recognized in the distribution (see text for references).
}
\label{fig:redshift}
\end{figure}

We plot in Fig. \ref{fig:redshift} the redshift distribution of the GOODS-ERS sample used in this work (black solid histogram) compared to that of the  GOODS-MUSIC sample adopted by \cite{fontana06} (red dotted histogram). Since the area covered by the ERS survey is relatively small, the sample is more sensitive to overdensities. The extended overdensities at $z\simeq 0.7$ and $z\simeq 1$, which cover the entire GOODS-S field (\citealt{vanzella05}, \citealt{salimbeni09} and references therein), are clearly recognizable. Unfortunately, the northern part of GOODS-S also includes a cluster at $z\simeq 1.6$ \citep{castellano07} and various groups at $z\simeq 2.2-2.3$ \citep{salimbeni09,yang10,magliocchetti11}, which both affect the overall redshift distribution.

Another difference from the \cite{fontana06} analysis is that the final sample used in this work includes Type 2 AGNs, since we show in \cite{santini12} that their stellar mass estimate is insignificantly affected by the nuclear emission. The same is untrue for Type 1 AGNs \citep{santini12}, so we removed spectroscopically identified Type 1 AGNs from the sample. Since their number is very small (only four sources identified in the entire sample), their removal does not affect the GSMF estimate. We also removed all identified Galactic stars. 
Finally, we applied a redshift selection to the range $0.6-4.5$ and we ended up with a sample of  2709 objects (of which 354 have spectroscopic redshifts).

\section{The galaxy stellar mass function (GSMF)} \label{sec:mf}

\subsection{Stellar masses}

Stellar masses were estimated by fitting the 14 band photometry (up to the 5.5 \mics rest-frame) to the Bruzual \& Charlot synthetic models, in both  their 2003 (BC03 hereafter) and 2007 version \citep[][CB07]{bruzual07}, through a $\chi^2$ minimization. 
For consistency with our previous works and  most of the studies in the literature, we adopted the estimates derived with BC03 templates as the reference ones. 
In the fitting procedure, redshifts were fixed to the spectroscopic or photometric ones. 
Our 1$\sigma$ errors, caused by both the photometric uncertainties and the photometric-redshift scatter, were computed by considering all the solutions within $\chi^2_{min}+1$. During the error computation, spectroscopic redshifts were fixed to their value, while photometric ones were allowed to vary  around their best-fit solution in order to account for their degeneracy.

\begin{figure}[!t]
\resizebox{\hsize}{!}{\includegraphics[angle=0]{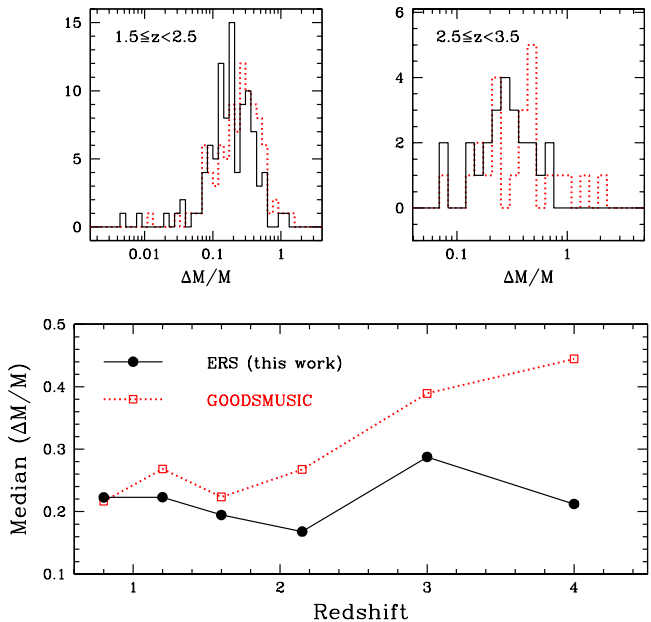}}
\caption{$Upper$ Distribution of the $\Delta M / M$ ratio, where $\Delta M$ is the average $1 \sigma$ error bar for each object, at $z\sim 2$ ($left$) and  $z\sim 3$ ($right$). Black solid histograms refer to the GOODS-ERS sample, whereas red dotted ones represent the GOODS-MUSIC data set. $Lower$ Median $\Delta M / M$ in each of the redshift bins used in this work as a function of the central redshift for GOODS-ERS (black solid circles/solid lines) and GOODS-MUSIC (red open boxes/dotted lines). 
}
\label{fig:errmass}
\end{figure}

We  parametrized the star formation histories as exponentially declining laws with a timescale $\tau$. We analysed a wide parameter space for metallicities, ages, extinctions, and $\tau$, whose details can be found in \cite{fontana04}, as updated  in \cite{santini09}. With respect to our previous works, we also excluded templates with super-solar metallicity at $z \geq 1$. Studies of the mass-metallicity relation \citep{maiolino08}  indeed demonstrated that galaxies at high redshift are typically characterized by sub-solar metallicities. 
We decided to adopt exponentially declining $\tau$ models despite it being likely that they are a poor and oversimplified description for the star formation history \citep[e.g.,][]{maraston10}. However, \cite{lee10}  showed that the resulting stellar masses can still be considered robust because of a combination of effects in the estimate of the galaxy star formation rates and ages. Moreover, $\tau$-models are widely used even in the most recent literature and allow a direct comparison with previous works.

We adopted a Salpeter IMF. We also computed the stellar masses by assuming a Chabrier IMF, and  checked that these are simply shifted by a factor -0.24 dex, which is constant to within 3\% at the different redshifts. Moreover, we tested that the GSMFs obtained by adopting the two IMFs are consistent after applying the same shift, in agreement with what was found by \cite{salimbeni09a}.  

The comparison with our previous GOODS-MUSIC sample allows us to test the effect of a deeper photometry data set on the accuracy of photometric redshifts and stellar masses. For this reason, we compared photometric redshifts and stellar masses for identical objects. 
The photometric redshifts of the present GOODS-ERS data are in very good agreement with the GOODS-MUSIC ones. Considering all objects in common between the two catalogues, the average scatter is $<|z_{ERS}-z_{GOODS-MUSIC}|/(1 + z_{ERS}) > = 0.07$, with only 0.06\% of severe (scatter $>0.5$) outliers. The stellar masses are also consistent with those derived from the GOODS-MUSIC catalogue. When selecting  galaxies  for which the redshift estimate differs by 0.1 at most, the scatter $(M_{ERS}-M_{GOODS-MUSIC})/M_{ERS}$ is on average equal to $-0.03 \pm 0.40$. 
The major improvement provided by the higher quality photometry of WFC3 observations leads to a reduction in the uncertainties in the stellar masses.  
In Fig. \ref{fig:errmass}, we compare the relative error in the stellar mass estimated using the GOODS-ERS data set (black solid circles/solid lines) with that obtained from GOODS-MUSIC catalogue (red open boxes/dotted lines). 
We again selected only galaxies common to both catalogues with consistent redshifts. 
In the upper panels, we show the distribution of the $\Delta M / M$ ratio, where $\Delta M$ is the average $1 \sigma$ error bar for each object ($\Delta M=(M_{*\ max}-M_{*\ min})/2$), in two redshift intervals centred at $z=2$ and $z=3$;  in the lower panel, we have plotted, as a function of redshift, the median $\Delta M / M$ in each of the redshift bins used in this work. 
The relative errors in the stellar mass for the GOODS-ERS sample are on average  $\sim 30\%$ smaller than those computed for the same objects using the GOODS-MUSIC photometry, the difference increasing with redshift. 
It is clear that deep photometry in the near-IR regime is crucial to help improve our stellar mass estimates.

\begin{figure*}[!t]
\resizebox{\hsize}{!}{\includegraphics[angle=270]{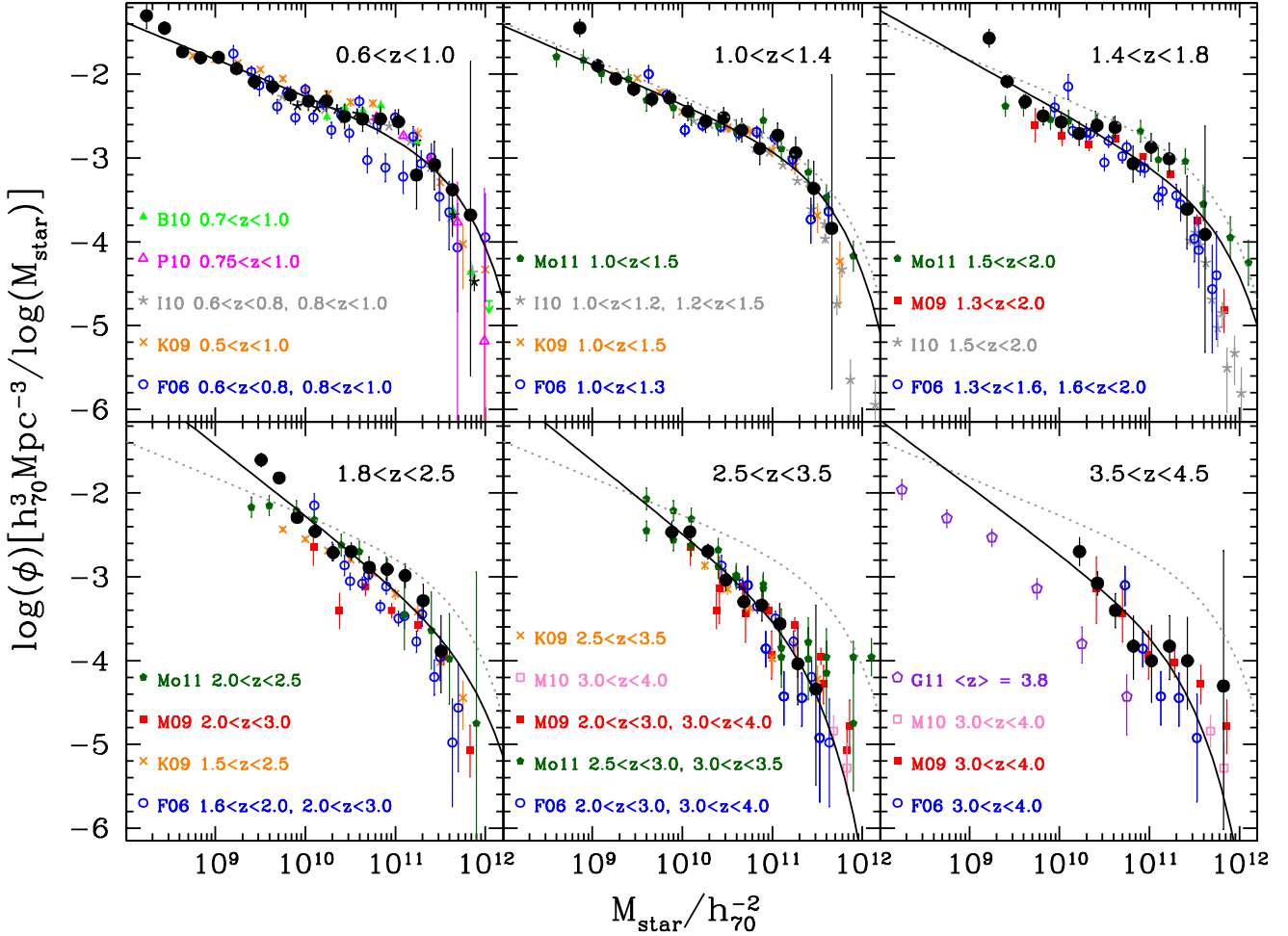}}
\caption{GSMFs obtained with BC03 stellar templates in different redshift ranges compared with previous works. Black solid circles represent our analysis with the $1/V_{max}$ method, black solid lines show the best-fit to a Schechter function according to the STY approach. The grey dotted line replicates the best-fit Schechter function at $z\sim 0.8$ in the higher redshift panels. Error bars include the uncertainties in the stellar masses as well as Poissonian errors. 
The highest mass points are often poorly determined, because of the large statistical error, resulting  from our poor sampling of the massive side. 
Other symbols represent  the $1/V_{max}$ results of previous works, scaled to the same cosmology and converted to the same IMF:  \cite{fontana06}: open blue circles (F06); \cite{bolzonella10}: solid green triangles (B10); \cite{pozzetti10}: open purple triangles (P10); \cite{ilbert10}: grey stars (I10); \cite{kajisawa09}: orange crosses (K09); \cite{marchesini09}: solid red boxes (M09); \cite{marchesini10}: open pink boxes (M10); 
\cite{mortlock11}: solid dark green pentagons (Mo11); \cite{gonzalez11}: open violet pentagons (G11). 
All the literature works considered for the comparison adopted the same stellar templates as this study. The legend shows the redshift intervals in which each set of points was computed.
}
\label{fig:mf_obs}
\end{figure*}

\subsection{The GSMF estimate} \label{sec:gsmf}

We estimated the GSMF by adopting both the non-parametric $1/V_{max}$ method \citep{schmidt68} and the STY \citep{sandage79} maximum-likelihood analysis assuming a Schechter parametric form.  As for any other magnitude-limited sample, our sample does not have a defined limit in stellar mass. For this reason, at each stellar mass and each redshift, we computed the fraction of objects lost because of the limited width of the $M_*/L$ distribution by adopting the technique described in \cite{fontana04}, after verifying that the simple parametrization used to describe the observed $M_*/L$ distribution still holds for our sample.

We show the results of our analysis as black solid circles ($1/V_{max}$ method) and black solid lines (STY approach) in Fig. \ref{fig:mf_obs} (where the reference BC03 templates were used). 
The best-fit Schechter parameters are reported in Tab. \ref{tab:parambc03}. 
The error bars in the $1/V_{max}$ points include Poissonian uncertainties, as well as uncertainties in the stellar masses. The latter were estimated by means of a Monte Carlo simulation, where we randomly extracted the stellar masses according to their $1 \sigma$ uncertainties and re-computed the GSMF of 10000 mock catalogues using the same procedure described above.

Given the degeneracy between the faint-end slope $\alpha$ and the characteristic mass $M^*$ of the Schechter function, the STY approach suffers from the incomplete sampling of the high mass regime owing to our small area, especially at high redshift. 
For this reason, the highest mass $1/V_{max}$ points are often poorly determined, because of the large statistical error. 
In the highest redshift bin, to constrain the fit, we fixed\footnote{In  Sect. \ref{sec:faintend}, we study how the best-fit parameter $\alpha$ varies when choosing a different value of $M^*$.} $M^*$ to the value obtained at $z\sim 3$. 
The fits found in this case are shown in Fig. \ref{fig:mf_obs} and the relevant Schechter parameters are given in Tab. \ref{tab:parambc03}. 
In this table, we also provide (third column) the fraction of objects where 
a secondary photometric-redshift solution falls outside each redshift interval. This fraction was defined following the criterion discussed above and allowing a tolerance in photometric redshift of 0.2. 
We found that the number density,  given by the normalization parameter $\phi^*$,  decreases with increasing redshift from $10^{-3.70^{+0.06}_{-0.07}}$ at $z\sim 0.8$ to $10^{-4.12^{+0.08}_{-0.10}}$ at $z\sim 4$. 
We recall that cosmic variance effects could cause oscillations in the normalization parameter, especially in the two bins that are most affected by the presence of overdensities, namely the $1.4-1.8$ and $1.8-2.5$ redshift intervals. However, a similarly decreasing trend for the normalization was also  observed by previous works \citep[e.g.][]{fontana06,perezgonzalez08,kajisawa09,marchesini09,mortlock11}.  
Most interestingly, the low-mass slope steepens significantly from $z\sim 0.8$ to $z\sim 3$, where the Schechter parameter $\alpha$ decreases from $-1.44 \pm 0.03$ to $-1.86 \pm 0.16$, and then flattens from $z\sim 3$ to $z\sim 4$. As demonstrated in Sect. \ref{sec:faintend}, this result remains valid here despite the uncertainties derived for the small area covered by our survey and the presence of known overdensities. Indeed, even the redshift ranges that are the most contaminated show  faint-end slopes in line with the results in the other redshift bins.

Up to $z\sim 2.5$, the GSMF shows a dip at $M_\star \simeq 10^{10} M_\odot$, which seems to shift to higher stellar masses as redshift increases, implying that a single Schechter is a poor parametrization. 
This dip has been identified in previous wide-field surveys and interpreted as  the differential evolution of the red and the blue populations \citep{ilbert10,pozzetti10,bolzonella10,mortlock11}.  
The effect is larger in the redshift intervals $1.4<z<1.8$ and $1.8<z<2.5$, which are highly affected by the presence of a well-known cluster at $z\sim 1.6$ \citep[e.g.][]{castellano07} and of several of localized overdensities at $z\simeq 2.2-2.3$ \citep{salimbeni09,yang10}, respectively: they are indeed populated by a higher fraction of old red galaxies, which enhances this dip. 
A different explanation of the dip around $\sim 10^{10} M_\odot$ was suggested by \cite{drory09}, who also  measured a bimodal shape in the GSMF of the blue and red populations separately. This dichotomy in galaxy formation, which pre-dates the red sequence appearance, was ascribed to a change with stellar mass in either star formation efficiency or galaxy assembly rate. 
The studies cited above show that a double Schechter is a more accurate description of the shape of the total GSMF. However, given the small size of our sample, the inclusion of two more free parameters \citep{bolzonella10} makes the fit degenerate and was not an approach that we adopted here.

\subsection{Comparison with previous results} \label{sec:comparison}

We show in Fig. \ref{fig:mf_obs} a compilation of $1/V_{max}$ points collected from the literature, as listed in the legend, scaled to the same cosmology and IMF. 
Unfortunately, it is impossible to correct for the effects of different stellar libraries, because, as we show in the next section, we are unable to determine a systematic shift that they could cause in the GSMF. Therefore, we decided to show only those studies that adopt the same stellar library as this work. 
Overall, our results are in good agreement with most of the other surveys, especially up to $z\sim 3$. 
In the two redshift intervals affected by the overdensities ($z\sim 1.6$ and $z\sim 2.2$), our GSMFs are on average higher than the other surveys, but still consistent with most of them within the errors.  
In general, we report a larger number of galaxies at the bright tail than \cite{fontana06}, because the present study includes AGNs (except the few identified Type 1), which preferentially live in high mass galaxies \citep{bundy08,alonsoherrero08,brusa09,silverman09,xue10,santini12}. 
Given the very deep near-IR observations used in this work, the sampling of the low-mass end of the GSMF is considerably finer than most previous surveys, on average by 0.5 dex up to $z\sim 1.8$  and by 0.1 dex at $z\sim 4$, and at the same time the conservative photometric cut ($K_S<25.5$) ensures reliable results even at the lowest masses. 

The only comparable study sampling similar or slightly lower stellar masses is the one of \cite{mortlock11}. This work is somewhat peculiar, being obtained from a set of biased pointings specifically designed to contain as many massive galaxies as possible, and a posteriori corrected to account for this bias. They  pushed their detection to $H=26.8$ at a $5\sigma$ level, while our sample, although extracted from images of similar depth, was cut at a brighter limit  to ensure good photometric quality. 
They also did not include any $K$ band data, which is important to estimate reliable stellar masses. Finally, since our study is based on 14 bands of photometry \citep[instead of 6 bands as][]{mortlock11}, our work also relies on good quality photometric redshifts.  

Despite the limited sky area, the bright-end tail is comparable overall within the uncertainties with that inferred by large surveys over the whole redshift range (with the  exception of the $1.4-2.5$ redshift interval, which, as discussed above, is affected by the presence of overdensities).  
The only severe disagreement is found when comparing our results in the highest redshift interval to \cite{gonzalez11}, who, as already pointed out in the introduction, derived the GSMF by using a different procedure, i.e.  by combining the UV luminosity function with an average $M_*/L$ ratio.

\subsection{The effect of different stellar templates} \label{sec:templates}

The systematic uncertainties caused by the various assumptions involved in spectral energy distribution   modelling were shown to dominate the overall error budget affecting the GSMF \citep[see][for a detailed analysis]{marchesini09}. In this regard, a significant role is played by the choice of the stellar templates used to estimate the stellar mass.

\begin{figure}[!t]
\resizebox{\hsize}{!}{\includegraphics[angle=0]{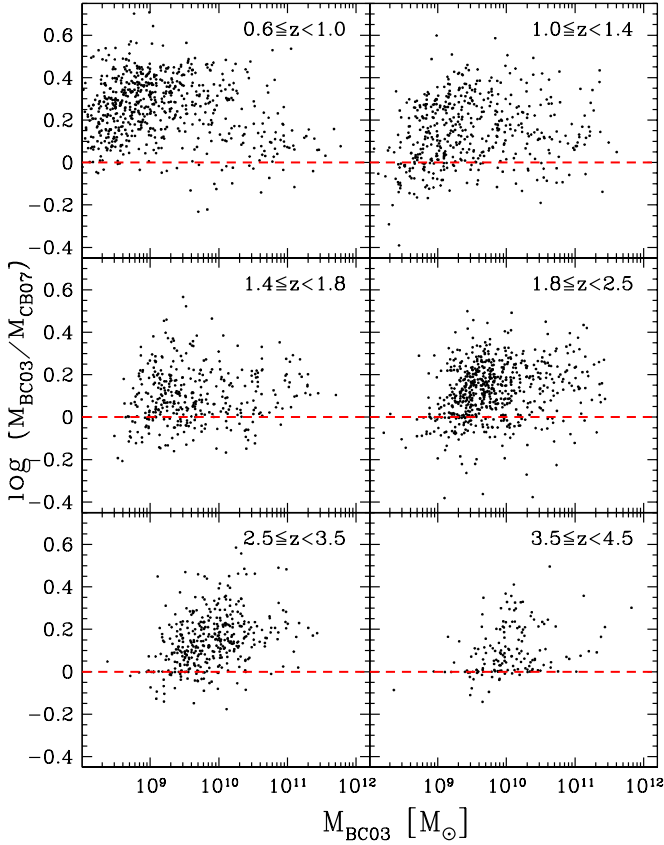}}
\caption{Ratio of stellar masses computed with BC03 ($M_{BC03}$) and CB07 ($M_{CB07}$)  templates versus $M_{BC03}$ in different redshift bins.
}
\label{fig:cfrmass}
\end{figure}

Stellar masses obtained using the CB07 stellar library, which includes an improved TP-AGB stars treatment, are on average 0.12 dex lower than those inferred using the BC03 templates, with a scatter as large as 0.17 dex.  
We plot in Fig. \ref{fig:cfrmass} their  ratio  as a function of the stellar mass adopted as a reference in this work ($M_{BC03}$) in different redshift bins. 
The lack of a clear trend of $M_{BC03}/M_{CB07}$ with stellar mass or redshift translates into a lack of a rigid offset between the GSMFs computed with the two libraries, although the CB07 points are on average at lower stellar masses than BC03.

\begin{figure}[!t]
\resizebox{\hsize}{!}{\includegraphics[angle=0]{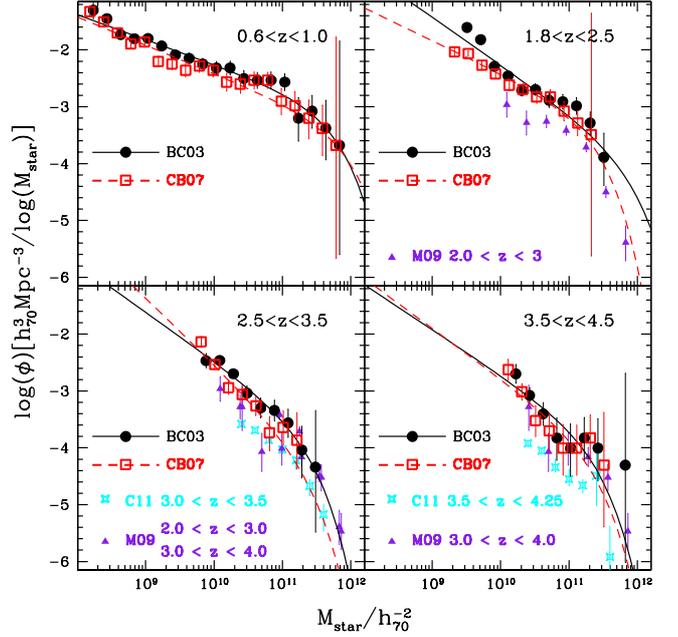}}
\caption{Comparison between the GSMFs obtained with the \cite{bc03} template library (black solid circles and solid curves) and  the \cite{bruzual07} one (red open boxes and dashed curves). Symbols are the results of the $1/V_{max}$ analysis and curves represent the STY Schechter fits. Error bars include the uncertainties in the stellar masses as well as Poissonian errors. 
Other symbols present $1/V_{max}$ results of previous works based on CB07 templates, scaled to the same cosmology and converted to the same IMF:  \cite{marchesini09}: solid purple triangles (M09); \cite{caputi11}: cyan stars. 
}
\label{fig:bc03cb07}
\end{figure}

We compare in Fig. \ref{fig:bc03cb07} the GSMFs obtained with the BC03 templates  (black solid curves/solid circles) and the CB07 ones (red dotted curves/open boxes).  For the sake of simplicity, we decided to report the four most representative bins. The results for the $1.0-1.4$ and $1.4-1.8$ redshift bins are very similar to the $0.6-1.0$ and $1.8-2.5$ ones, respectively. 
We also show the $1/V_{max}$ points of \cite{marchesini09} (their set 8) and \cite{caputi11}, both obtained by adopting the CB07 templates. 
The results of \cite{marchesini09} agree with our CB07-based GSMF in all except  the $1.8-2.5$ redshift interval, likely because of imperfect redshift overlap between the two analysis. 
The points from \cite{caputi11} are in broad agreement with ours at the bright end,  while the incompleteness that the authors claim to be affected by below $M_*\sim 10^{11} M_\odot$  is likely responsible for the disagreement at low stellar masses. 

The best-fit Schechter parameters of the CB07-based GSMF are reported in  Tab. \ref{tab:paramcb07}. 
At $z>2.5$, we were forced to fix\footnote{See Sect. \ref{sec:faintend} for an analysis of how the best-fit parameter $\alpha$ varies when choosing a different value for $M^*$.} the $M^*$ parameter to its best-fit value at $z\sim 2.15$. If it is instead allowed to vary, the fit is unconstrained or the maximum-likelihood analysis does not converge.  
The CB07- and BC03-based GSMFs differ from each other. 
However, we do not find any similar systematic behaviour at all redshifts. 
That the high-mass end of the CB07-based GSMF is unconstrained at $z>2.5$, while the BC03-based one suffers from poor statistical sampling only in the highest redshift bin ($z>3.5$), is a further confirmation that the two GSMFs are not affected by a systematic shift in stellar mass. 

At the lowest and the highest redshifts, we find the closer agreement, the normalization of the best-fit Schechter function being only slightly lower when CB07 templates are used. 
At intermediate redshifts, we observe a more serious disagreement, resulting in different faint-end slopes and characteristic masses. 
This is unsurprising, because the effect of  including of the TP-AGB phase is expected to be important at intermediate ages ($0.2 - 2$ Gyr), which predominate the $2 \lesssim z \lesssim 3$ redshift range \citep{maraston05,henriques11}. 
Although the difference between the CB07- and the BC03-based GSMFs do not show a systematic trend at all redshifts, the characteristic masses seem to be on average lower when CB07 templates are used (see Fig. \ref{fig:contours} discussed in the next section), as expected, despite the large uncertainties, in agreement with the results of \cite{marchesini09} . 
This trend is clear at $1.4 < z < 2.5$, where our redshift bins overlap with those of \cite{marchesini09}, while the lack of high quality statistics at higher redshifts prevents us from drawing any firm conclusions about the effect of changing the stellar templates. For what concerns the variation in $\alpha$ when changing the template library, we find similar slopes from $z\sim 0.8$ to $z\sim 1.2$, while in the redshift interval $1.4- 2.5$ the BC03-based GSMFs are steeper than the CB07-based ones by $0.2-0.3$. 
\cite{marchesini09} reports similar slopes when using BC03 and CB07 stellar templates in the redshift interval $1.3 - 3.0$, while their BC03-based GSMF is steeper than the CB07-based one at $3.0<z<4.0$. 
The intrinsic difference between the two surveys does not allow us to investigate the origin of this mismatch.

\section{The faint-end slope} \label{sec:faintend}

The main goal of this study has been to investigate the faint-end slope of the GSMF, especially at the highest redshifts ($z>2$). From both  Fig. \ref{fig:mf_obs} and Tables  \ref{tab:parambc03} and \ref{tab:paramcb07}, it is evident that the low-mass tail  steepens with redshift. 
The results from  applying the STY approach to our BC03-based data indicate that the faint-end slope steepens significantly between $z\sim 0.8$, where we fitted $\alpha = -1.44 \pm 0.03$, and $z\sim 3$, where the best-fit $\alpha$ is equal to $-1.86 \pm 0.16$, before flattening up to $z\sim 4$.  

First of all, we performed a simple sanity check to verify that the abundance of low mass objects at $z > 1.8$ is reliable by plotting all objects with $M_* < 10^{10}M_\odot$ and $1.8<z<2.5$ on a $BzK$ diagram. For galaxies at $z>2.5$, we adopted the analogous $RJL$ diagram (using  IRAC 3.6 \mics as $L$ band), which extends the former to the $2.5<z<4$ redshift regime \citep{daddi04}, and checked stellar masses below $ 2 \cdot 10^{10}M_\odot$. 
Approximately 91\% of the sources indeed lie in the high redshift regions of these diagrams, making us confident of their photometric redshift estimate.
As an additional check, we carefully inspected the individual photometric-redshift probability distribution curves for each source with $z> 1.8$ and $M_* <2 \cdot 10^{10}M_\odot$. Following the criterion described in Sect. \ref{sec:data}, we found that 96.5\% of these sources have a 90\% probability of lying at $z>1.5$.

\begin{figure}[!t]
\resizebox{\hsize}{!}{\includegraphics[angle=0]{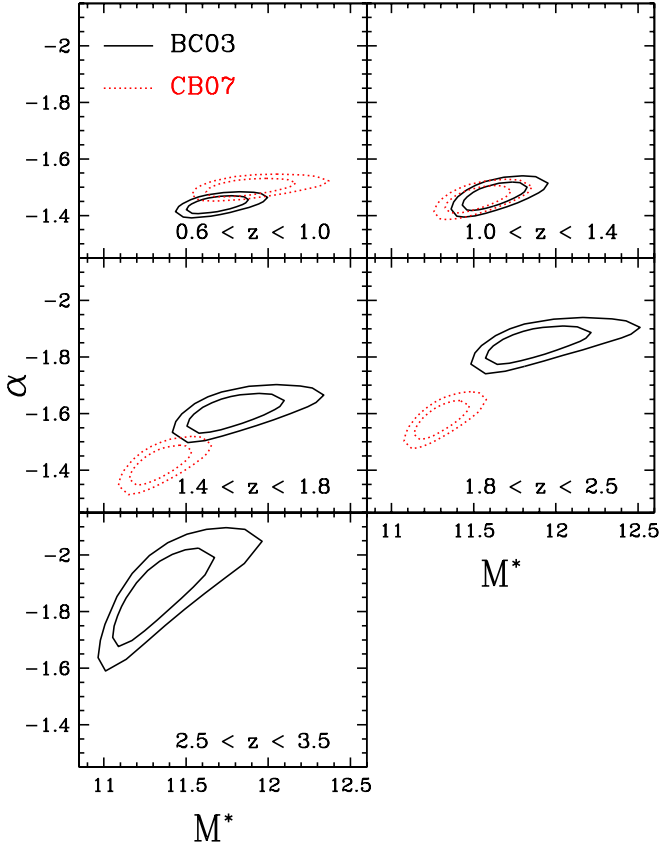}}
\caption{The $\alpha$ and $M^*$  space ($1\sigma$ and $2\sigma$ contours) resulting from the maximum-likelihood analysis.  Black solid curves refer to BC03-based GSMFs, red dotted curves refer to CB07-based ones. }
\label{fig:contours}
\end{figure}

\begin{figure}[!t]
\resizebox{\hsize}{!}{\includegraphics[angle=0]{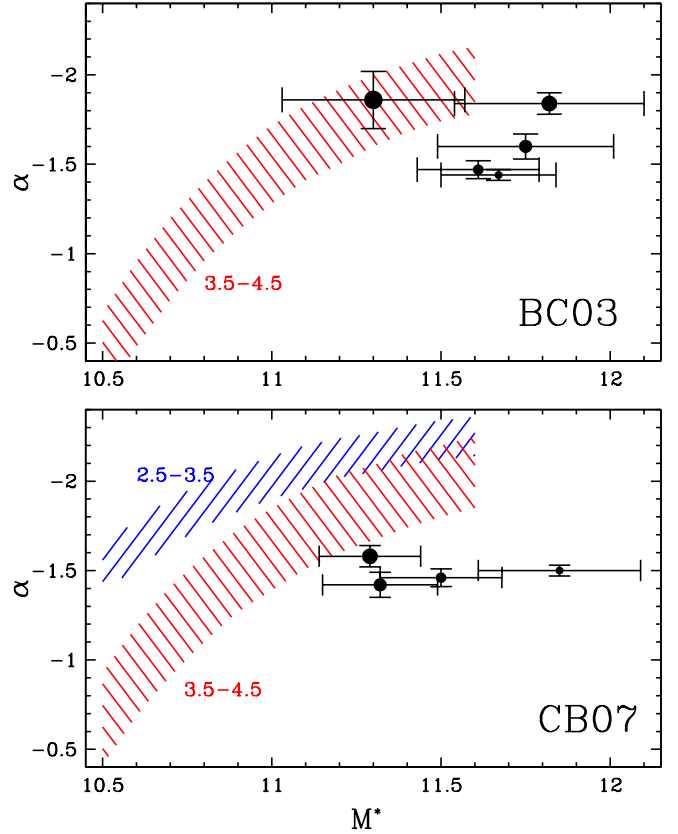}}
\caption{
Best-fit $M^*$ and $\alpha$ model parameters obtained by adopting BC03  ($upper$ panel) and CB07 ($lower$ panel) templates at different redshifts. The symbol size increases with redshift. The shaded areas show the values of the faint-end slope at different fixed $M^*$ in the redshift bins indicated by the labels. }
\label{fig:mfix}
\end{figure}

As already pointed out in Sect. \ref{sec:gsmf}, the small sky area sampled by our data may be responsible for degeneracies between the faint-end slope $\alpha$ and the characteristic mass $M^*$ when fitting a Schechter function. 
We therefore studied in detail the degeneracies in the $\alpha - M^*$  plane. 
The results are shown in Figs. \ref{fig:contours} and \ref{fig:mfix}. In the first figure, we analysed the redshift intervals where both parameters were allowed to vary, while in the second one we studied the dependence of the best-fit $\alpha$ on the chosen $M^*$ in those redshift bins where we were forced to fix the characteristic mass  to constrain the maximum-likelihood analysis. 
 
In Fig. \ref{fig:contours}, we show the $1\sigma$ and $2\sigma$ contours for $\alpha$ and $M^*$ Schechter parameters, for both the BC03-based GSMFs (black solid curves) and CB07-based ones (red dotted curves). 
While the parameter $\alpha$ is well-constrained at all redshifts (although with uncertainties increasing with $z$), our data prevent us from properly inferring the value of the characteristic mass. Nonetheless, as we show below, the result on $\alpha$ is robust against the degeneracy of $M^*$. 

The steepening in $\alpha$ between $z\sim 0.8$ and  $z\sim 3$  is clear from Fig. \ref{fig:contours} when the BC03 stellar library is used. When we instead adopted CB07 templates, the faint-end slope did not change much from $z\sim 0.8$  to $z\sim 2.2$, while at higher redshifts ($2.5<z<3.5$) we were forced to fix the value of $M^*$ to constrain the fit (see Sect. \ref{sec:templates}), making the best-fit $\alpha$ parameter dependent on the choice of the characteristic mass.

To study how the best-fit value of $\alpha$ changes when varying the $M^*$ value, we built a grid of log($M^*[M_\odot]$) ranging from 10.5 to 11.6 (these limits are justified by previous works) with steps in mass of 0.1 and we fitted a Schechter function to the data for each element of the grid. We adopted this procedure in all the redshifts bins where the fit is unconstrained. We show in Fig. \ref{fig:mfix} the $\alpha - M^*$ plane, where the shaded areas show the values of the faint-end slope at different fixed $M^*$ in the redshift bins indicated by the labels. The symbols show the best-fit values (and their uncertainties) for $\alpha$ and $M^*$ at $z<3.5$ ($z<2.5$ when using CB07 templates), where our results are insignificantly affected by the lack of high quality statistics and we could allow both parameters to vary.  The upper panel refers to the BC03-based GSMFs, while the lower one is obtained by adopting CB07 stellar templates.

From Fig.\ref{fig:mfix} (lower panel, blue shaded region), it appears that, whatever reasonable value for $M^*$ is chosen at $2.5<z<3.5$, the best-fit $\alpha$ is clearly steeper than the best-fit values at lower redshifts, confirming the result found with BC03 templates and supporting that the major result of this paper is unaffected by the poor sampling of the high mass regime. 
However, as shown by the red shaded regions in Fig. \ref{fig:mfix},  presently available data do not allow us to draw any firm conclusion about the trend of $\alpha$ between $z\sim 3$ and $z\sim 4$.

To  corroborate our result for the slope of the low-mass end, and ensure that is unaffected by the lack of statistics at the massive end, we took advantage of the outcomes of large surveys and followed different approaches. 
We fitted the  $1/V_{max}$ points from this study together with those collected from the literature in comparable redshift intervals.  We note that in principle, and in contrast to  the STY approach, fitting $1/V_{max}$ points involves data binning, thus may in general produce a different fit. 
We included only those surveys whose results are obtained using a method similar to our own and that sample the high-mass tail of the distribution, typically above $M_*\simeq 3 \times 10^{10}M_\odot$.  
However, we obtained very similar results when also including  the points from the literature at lower masses.   
We found that a single Schechter function does not seem to reproduce the faint- and the bright-end simultaneously in a satisfactory way. 
This is unsurprising because the Schechter function is itself a poor description of the shape of the GSMF  when samples with high quality statistics are used (see discussion in the introduction and in Sect. \ref{sec:gsmf}). 
However, the inhomogeneity of the data set can also play a role: we collected $1/V_{max}$ points from different surveys, observed in different sky areas, and computed with slightly different methods. 
We then fitted the ensemble of the $1/V_{max}$ points from this work plus those collected from the literature with a double power-law\footnote{The assumed functional shape is 
$\phi_* /[(M/M^*)^{-(1+\alpha)}+(M/M^*)^{-(1+\beta)}]$ }. The best-fit parameters are shown in Tab. \ref{tab:paramdpl}. This analytic shape, having one additional degree of freedom than a single Schechter function, provides a tighter fit to the data at all redshifts.

\begin{figure}[!t]
\resizebox{\hsize}{!}{\includegraphics[angle=270]{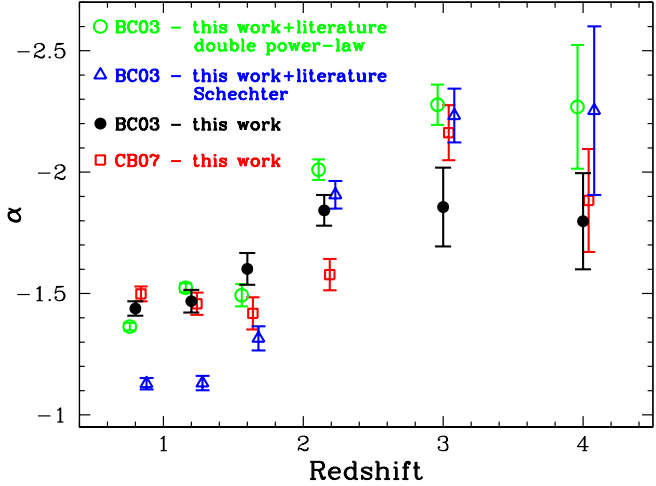}}
\caption{Faint-end slope as a function of redshift. The parameter $\alpha$ was computed through a maximum-likelihood analysis with a Schechter form (black solid circles refer to the BC03 library, red open boxes refer to the CB07 one) and by fitting the ensemble (this study + previous surveys) of $1/V_{max}$ points with a Schechter  parametric form (blue open triangles) and a double power-law shape (see text, green open circles). 
The different sets are shifted in redshift with respect to the central values in each interval (shown by the solid black circles).}
\label{fig:alpha}
\end{figure}

\begin{table*}
\centering
\begin{tabular} {ccccccc}
\hline \hline 
\multicolumn{6}{c}{Schechter parameters (STY method) -- BC03 templates}\\
\hline
\noalign{\smallskip} 
Redshift bin & N & \%$_{not\ secure}$ & $\alpha$ & log$_{10}$M$^*$(M$_\odot$) & log$_{10}\phi_*$(Mpc$^{-3}$) & log $\rho$(M$_\odot$Mpc$^{-3}$) \\
\noalign{\smallskip} \hline \noalign{\smallskip}
0.6 - 1.0 & 584 & 8.5\% & -1.44 $\pm$ 0.03 & 11.67 $\pm$ 0.17 & -3.70 $^{+0.06}_{-0.07}$ & 8.51 \\
1.0 - 1.4 & 375 & 6.2\% & -1.47 $\pm$ 0.05 & 11.61 $\pm$ 0.18 & -3.47 $^{+0.07}_{-0.08}$ & 8.35 \\
1.4 - 1.8 & 259 & 9.7\% & -1.60 $\pm$ 0.07 & 11.75 $\pm$ 0.26 & -3.85  $^{+0.11}_{-0.14}$& 8.23 \\
1.8 - 2.5 & 425 & 7.4\% & -1.84 $\pm$ 0.06 & 11.82 $\pm$ 0.28 & -4.17  $^{+0.14}_{-0.20}$& 8.29 \\
2.5 - 3.5 & 182 & 11.8\% & -1.86 $\pm$ 0.16 & 11.30 $\pm$ 0.27 & -3.94 $^{+0.16}_{-0.26}$& 7.97 \\
3.5 - 4.5 & 51 & 8.4\% & -1.80 $\pm$ 0.20 & 11.30 (fixed) & -4.12 $^{+0.08}_{-0.10}$ & 7.72 \\
\noalign{\smallskip} \hline \noalign{\smallskip}
\end{tabular}
\caption{Best-fit Schechter parameters in the different redshift intervals  as a result of the STY approach using \cite{bc03} templates. Parameters with no error bars have been fixed to the value in the lower redshift bin.  The second column indicates the numbers of galaxies in each redshift bin based on which the GSMF is actually computed. The third column shows the fraction of galaxies where a secondary redshift solution outside the redshift bin cannot be discarded with a 90\% probability (see Sect. \ref{sec:gsmf} for details).	
The last column reports the corresponding mass density $\rho$ obtained by integrating the GSMF between $10^8$ and $10^{13}$ M$_\odot$. 
}\label{tab:parambc03}
\end{table*}

\begin{table*}
\centering
\begin{tabular} {ccccccc}
\hline \hline 
\multicolumn{5}{c}{Schechter parameters (STY method) -- CB07 templates}\\
\hline
\noalign{\smallskip} 
Redshift bin & N & \%$_{not\ secure}$ & $\alpha$ & log$_{10}$M$^*$(M$_\odot$) & log$_{10}\phi_*$(Mpc$^{-3}$) & log $\rho$(M$_\odot$Mpc$^{-3}$) \\
\noalign{\smallskip} \hline \noalign{\smallskip}
0.6 - 1.0 & 509 & 8.5\% & -1.50 $\pm$ 0.03 & 11.85 $\pm$ 0.24 & -3.70 $^{+0.10}_{-0.13}$ & 8.39 \\
1.0 - 1.4 & 372 & 6.2\% & -1.46 $\pm$ 0.05 & 11.50 $\pm$ 0.18 & -3.49 $^{+0.08}_{-0.09}$ & 8.22 \\
1.4 - 1.8 & 264 & 9.7\% & -1.42 $\pm$ 0.07 & 11.32 $\pm$ 0.17 & -3.41  $^{+0.09}_{-0.12}$& 8.09 \\
1.8 - 2.5 & 437 & 7.4\% & -1.58 $\pm$ 0.06 & 11.29 $\pm$ 0.15 & -3.52  $^{+0.10}_{-0.12}$& 8.07 \\
2.5 - 3.5 & 153 & 11.8\% & -2.16 $\pm$ 0.11 & 11.29 (fixed) & -4.39 $^{+0.28}_{-1.14}$& 8.05 \\
3.5 - 4.5 & 45 & 8.4\% & -1.88 $\pm$ 0.21 & 11.29 (fixed) & -4.28 $^{+0.11}_{-0.15}$ & 7.67 \\
\noalign{\smallskip} \hline \noalign{\smallskip}
\end{tabular}
\caption{Same as Tab. \ref{tab:parambc03} using \cite{bruzual07} templates.  
}\label{tab:paramcb07}
\end{table*}

\begin{table*}
\centering
\begin{tabular} {cccccc}
\hline \hline 
\multicolumn{6}{c}{Double power-law parameters (fit to $1/V_{max}$ points) -- BC03 templates}\\
\hline
\noalign{\smallskip} 
Redshift bin & $\alpha$ & $\beta$ & log$_{10}$M$^*$(M$_\odot$) & log$_{10}\phi_*$(Mpc$^{-3}$) & log $\rho$(M$_\odot$Mpc$^{-3}$) \\
\noalign{\smallskip} \hline \noalign{\smallskip}
0.6 - 1.0 & -1.36 $\pm$ 0.02 & -4.47 $\pm$ 0.12 & 11.39 $\pm$ 0.01 & -2.75 $^{+0.02}_{-0.02}$ & 8.50 \\
1.0 - 1.4 & -1.52 $\pm$ 0.02 & -5.24 $\pm$ 0.16 & 11.38 $\pm$ 0.01 & -3.10 $^{+0.02}_{-0.02}$ & 8.24 \\
1.4 - 1.8 & -1.49 $\pm$ 0.05 & -4.70 $\pm$ 0.23 & 11.30 $\pm$ 0.03 & -3.12  $^{+0.05}_{-0.06}$& 8.12 \\
1.8 - 2.5 & -2.01 $\pm$ 0.04 & -6.25 $\pm$ 1.57 & 11.64 $\pm$ 0.06 & -3.94  $^{+0.09}_{-0.12}$& 8.28 \\
2.5 - 3.5 & -2.28 $\pm$ 0.08 & -6.70 $\pm$ 4.84 & 11.77 $\pm$ 0.10 & -4.73 $^{+0.17}_{-0.28}$& 8.24 \\
3.5 - 4.5 & -2.27 $\pm$ 0.25 & -6.38 $\pm$ 7.48 & 11.81 $\pm$ 0.19 & -4.84 $^{+0.34}_{-4.84}$ & 8.15 \\
\noalign{\smallskip} \hline \noalign{\smallskip}
\end{tabular}
\caption{Best-fit parameters of the double power-law shape fit to $1/V_{max}$ points from this work (using BC03 templates) + a collection from the literature at $M_* \gtrsim 3 \times 10^{10}M_\odot$ (see text). The last column reports the corresponding mass density $\rho$ obtained by integrating the GSMF between $10^8$ and $10^{13}$ M$_\odot$.  
}\label{tab:paramdpl}
\end{table*}

We report the different values of the faint-end slope as a function of redshift in Fig.  \ref{fig:alpha}. 
It is shown that, regardless of the stellar templates and method  adopted and the functional shape  fitted to the data, all the results indicate a steepening of the faint-end of the GSMF with redshift up to $z\sim 3$. The trend is robust despite the relatively large error bars, especially at high redshift, and the presence of known overdensities at $z\sim 1.6$ and $z\sim 2.2-2.3$, and it is unaffected by the lack of high quality statistics at the high-mass end typical of small sky areas. 
The steepening of the faint-end slope with redshift seems to halt at $z>3$ and the value of $\alpha$ seems to remain constant up to $z \sim 4$. However,  although this is confirmed by the use of the outcome of previous large surveys, the results based on our data alone are largely dependent on the choice of the fixed $M^*$ parameter.

The tendency for the low-mass end of the GSMF to steepen with redshift was previously  found by other authors. 
According to the evolutionary STY fit of \cite{fontana06}, $\alpha$ ranges from $-1.25\pm 0.03$ at $z\sim 0.8$ to $-1.51\pm 0.13$ at $z\sim 4$.  
Increasing with redshift but flatter values for the faint-end slope were  obtained by \cite{marchesini09}, who reported $\alpha = -0.99$ at $z\sim 1.6$ and $\alpha = -1.39$ at $z\sim 3.5$. \cite{kajisawa09} found $\alpha = -1.26^{+0.03}_{-0.03}$ at $z\sim 0.75$ and $\alpha = -1.75^{+0.15}_{-0.13}$ at $z\sim 3$. Very steep GSMFs, in agreement with those inferred with our data, were fitted by \cite{caputi11}, who measured $\alpha = -2.07^{+0.08}_{-0.07}$ at $z\sim 3.9$. Finally, \cite{mortlock11} reported $\alpha = -1.36\pm 0.05$ at $z\sim 1.25$ and  $\alpha = -1.89\pm 0.11$ at $z\sim 2.75$. Both \cite{mortlock11} and \cite{caputi11} found a flattening at higher redshift similar to the one that we measure at $z\sim 4$, i.e. they fitted  $\alpha = -1.73\pm 0.09$ at $z\sim 3.25$ and  $\alpha = -1.85^{+0.27}_{-0.32}$ at $z\sim 4.6$, respectively.

We note that  equally robust results cannot be inferred for the evolution of the characteristic mass $M^*$, whose best-fit values are highly sensitive to the stellar templates and  the functional shape fitted to the data (see Tabs. \ref{tab:parambc03}, \ref{tab:paramcb07} and \ref{tab:paramdpl}), as well as  the size of the sample.

\section{The stellar mass density (SMD)}\label{sec:md}

\begin{figure}[!t]
\resizebox{\hsize}{!}{\includegraphics[angle=0]{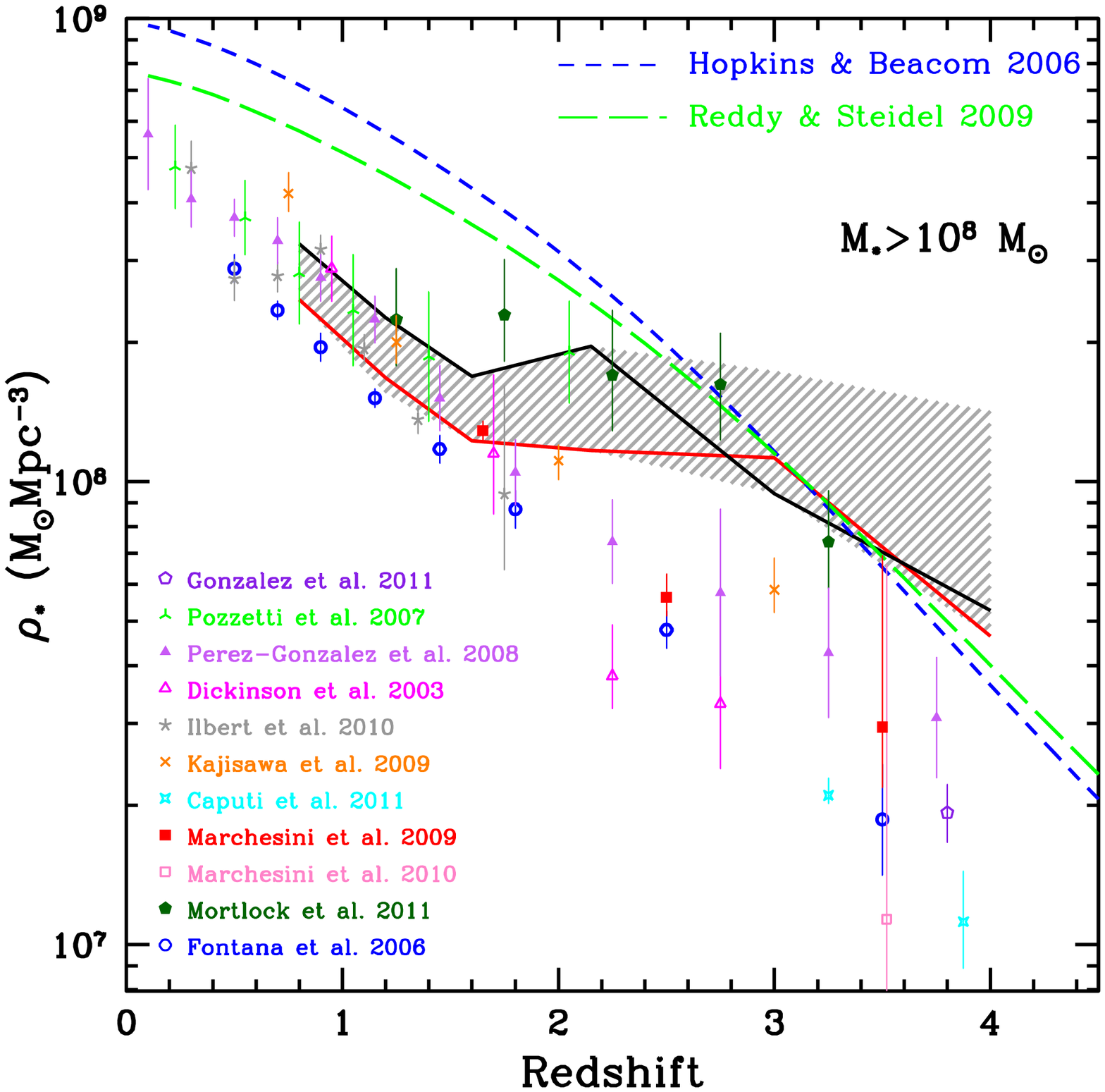}}
\caption{SMD between $10^8$ and $10^{13} M_\odot$ versus redshift. The solid black and red curves are the integral of the STY Schechter fits using BC03 and CB07 templates, respectively. 
The grey shaded area shows the dispersion obtained when integrating the fit, with both a Schechter and a double power-law functional shape, of our $1/V_{max}$ points together with those from the literature 
(see text and Fig. \ref{fig:alpha}). 
Coloured symbols represent a compilation of results from the literature as listed in the legend (see text for their integration limits). All the points are scaled to the same cosmology and IMF. All the results are based on BC03 stellar templates, except the \cite{caputi11} work, which adopts the CB07  library. 
As far as \cite{marchesini10} results are concerned, the central redshift was shifted by 0.02 for visualization purposes; given their Tab. 6, the central values is reported and the error bar indicates the total scatter in their estimates. 
The blue dashed and green long-dashed lines show the integrated star formation history according to the best-fit relation from \cite{hopkins06} and \cite{reddy09}, respectively. 
}
\label{fig:mdens}
\end{figure}

We computed the total SMD by integrating the  analytical fitting functions in each redshift bin from $10^8$ to $10^{13} M_\odot$. 
We show in Fig. \ref{fig:mdens} the SMD derived from the STY analysis using BC03 and CB07 templates as solid black and red curves, respectively. 
We also show a compilation of results from the literature, reported to the same cosmology and IMF, as listed in the legend. The same integration limits as in this study were used in most of the works considered. The only exceptions are the \cite{mortlock11} points ($M_*>10^7 M_\odot$), the \cite{ilbert10} ones ($M_*>10^5 M_\odot$), and those from \cite{dickinson03} and \cite{perezgonzalez08}, who adopted redshift-dependent mass limits (we refer to these works for more details). 
Our results show good agreement with those computed by previous authors at $0.6 \lesssim z \lesssim 2$, although we recall once again that our mass densities in the redshift intervals around $z\sim 1.6$ and $z\sim 2.2-2.3$ might be systematically too high owing to a few known overdensities. 
The steepness in the faint-end of the GSMF computed by this work is responsible for the large values of the SMD inferred at $z>2$. Our estimates are higher than those reported by previous authors, with the exception of the \cite{mortlock11} results. However, the latter results originate from a different shape of the GSMF: \cite{mortlock11} indeed found flatter faint-end slopes than we do, and the large SMD is a consequence of a higher density of high mass galaxies (see Fig. \ref{fig:mf_obs}).  

We note that the SMD is affected by uncertainties caused by systematic effects. In Fig. \ref{fig:mdens}, the grey shaded region  indicates the dispersion in the SMD 
when including the outputs obtained by integrating the fit to our $1/V_{max}$ points plus those collected from the literature with both a Schechter function and a double power-law shape (see Sect. \ref{sec:faintend}). 
This region represents the systematic errors caused by the choice of the stellar library and  the functional shape of the GSMF, as well as  the simultaneous use of the ERS observations as a probe of the low-mass end of the GSMF and the results of large surveys to constrain the bright-end.
The dispersion increases significantly at $z\gtrsim 3$, reflecting the large scatter among existing surveys.  Moreover, the lack of overlap with most previous results at these redshifts is a sign of the impossibility to assemble a single, self-consistent GSMF from the highest to the lowest masses.
This is due to the inhomogeneity of the samples, to the variance between different fields and also to the intrinsic uncertainties at high redshift in both the stellar masses and GSMF. 

We  compared the SMD with the integrated star formation rate density. 
For this purpose, we first considered the best-fit to the compilation of SFRD measurements made by \cite{hopkins06}. 
Following \cite{wilkins08}, we rescaled it to a Salpeter IMF and integrated it as a function of time, 
after accounting for the gas recycle fraction. 
The latter is the fraction of stellar mass  returned to the interstellar medium as a function of time, and was computed using the \cite{bc03} model. The result of this calculation is shown in Fig. \ref{fig:mdens} by the blue dashed line. 
We then performed the same calculation by using the best-fit parametric shape for the star formation history inferred by \cite{reddy09}, which also includes  more recent high redshift points as well as a luminosity-dependent dust correction to the $z>2$ data. We obtained the green long dashed line shown in Fig. \ref{fig:mdens}. 

Our results solve the discrepancy between the SMD and the integrated SFRD 
at $z>2$ (modulo the uncertainties affecting the $z\sim 2.1$ redshift interval), especially when considering the dispersion caused by the inclusion of high mass $1/V_{max}$ points from the other surveys. Consistency at high redshift was found by \cite{mortlock11} and  \cite{papovich11}, the latter study being based on an independents analysis. 
Overall, our results support the notion that the SMD can be reasonably close to the  integrated SFRD at $z>2$, mostly due to a steepening of the GSMF, although our results might be systematically too high because of the known overdensities in the small ERS field. 
As mentioned, the higher values that we obtained than most previous studies is essentially due to the efficiency of WFC3 deep near-IR data to accurately recover the faint-end of the GSMF, especially at high redshift, which contributes significantly to the total SMD. 
However, the significant steepening in the faint-end slope presented in this work is insufficient to solve the disagreement at $z<2$, where the integrated SFRD exceeds the observed SMD by a factor of $\sim 2-3$, even when both of them are integrated down to low values of stellar mass / luminosity  and  adopting the SFRD computed by \cite{reddy09}. 
Their SFRD, although it is 
lower than that resulting from the best-fit relation of \cite{hopkins06} at $z<2.5$, is still unable to reconcile the two observables. 
The discrepancy is also  not solved when our deep data, which allow a good control of the faint-end slope, are matched to the large surveys results to constrain the bright tail of the GSMF, 
and it gets even worse if one assumes that our mass density results are systematically too  high owing to overdensities in the ERS field.

\section{Comparison with theoretical models}\label{sec:models}

\begin{figure}[!t]
\resizebox{\hsize}{!}{\includegraphics[angle=0]{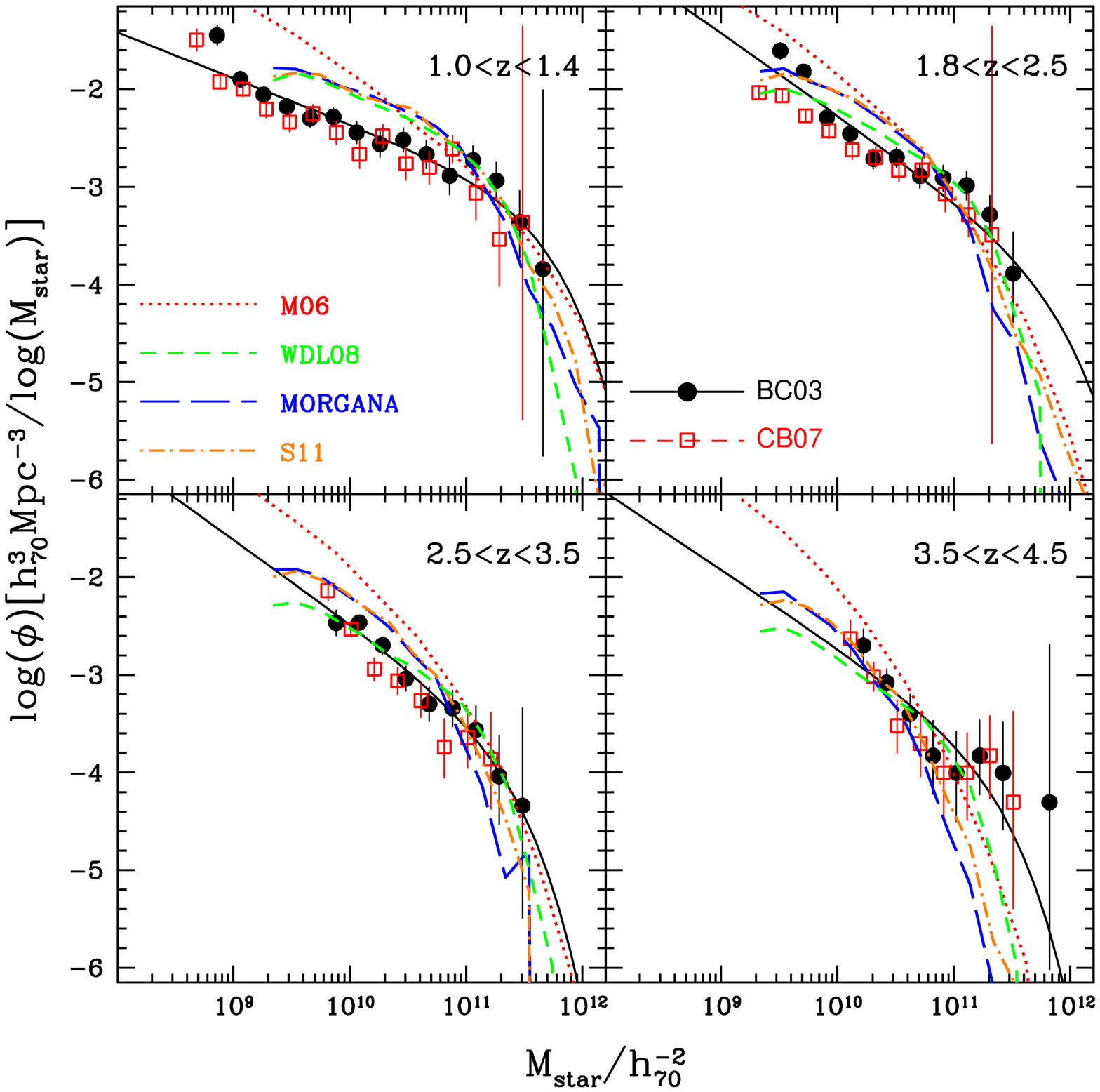}}
\resizebox{\hsize}{!}{\includegraphics[angle=0]{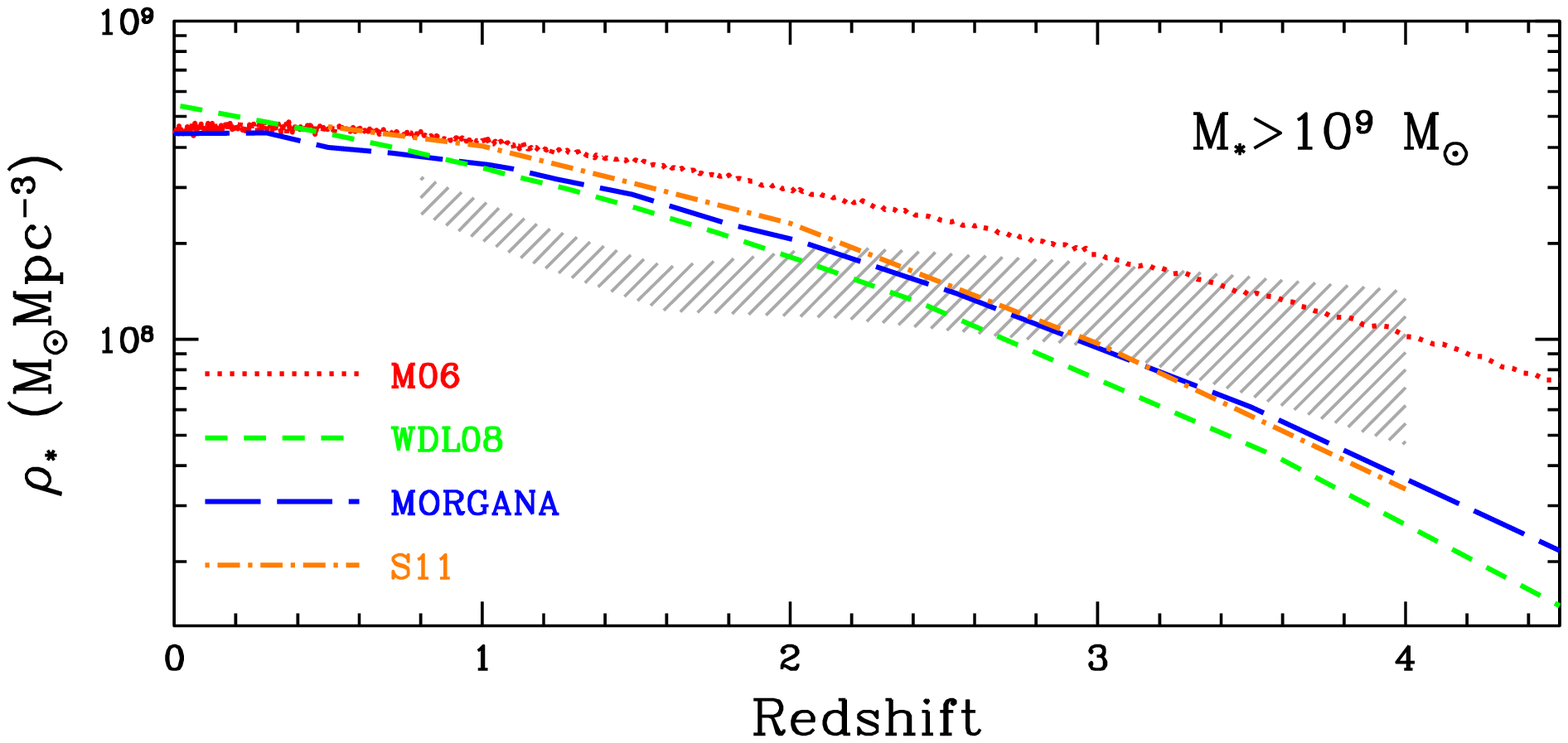}}
\caption{Observed GSMFs in different redshift ranges
  ($upper$) and observed SMD between $10^9$ and
  $10^{13} M_\odot$ ($lower$) compared with theoretical predictions,
  shown as coloured curves.  Red dotted curves: \cite{menci06}. Blue
  long-dashed curves: MORGANA \citep{monaco07}. Green dashed curves:
  \cite{wang08}. Orange dot-dashed curves: \cite{somerville11}.  Black
  solid circles, solid lines, red open boxes, and the grey shaded area show the
  results of the present work as in Figs. \ref{fig:bc03cb07} and
  \ref{fig:mdens}.  }
\label{fig:mf_mod}
\end{figure}

In Fig. \ref{fig:mf_mod}, we compare our results with
the predictions of semi-analytical models of galaxy formation and
evolution, which follow the evolution of the baryonic component adopting  an approximate description of the relevant physical processes
(i.e. gas cooling, star formation, stellar feedback, black hole growth,
and AGN feedback) and of their interplay with gravitational processes,
linked to the assembly of the large-scale structure of the
Universe. These ``recipes'' include a number of parameters that are
usually fixed by comparing model predictions with a set of
low-redshift observations. Despite their simplified approach, semi-analytical models
have turned into a flexible and widely used tool to explore a broad
range of specific physical assumptions, as well as the interplay
between different physical processes. 
We considered different, independently developed semi-analytical models:
\cite{menci06} (red dotted curves) updated to include the
\cite{reed07} halo mass function, MORGANA (\citealt{monaco07}, as
updated in \citealt{lofaro09}, blue long-dashed curves), \cite{wang08} (green
  dashed curves), and \cite{somerville11} (orange dot-dashed
  curves).  We refer to the original papers for a
detailed description of the recipes adopted in the galaxy formation
and evolution modelling. All three models of \cite{wang08}, \cite{somerville11}, and MORGANA resolve galaxies with $M_* > 10^{9} M_\odot$, while the \cite{menci06} model has a lower mass limit of $10^8 M_\odot$.  
The predicted stellar masses were convolved with a Gaussian
error distribution on log$M_*$ (see Fig. \ref{fig:errmass}) 
 to reproduce the observational uncertainties. We refer the
reader to \cite{fontanot09} and \cite{marchesini09} for a detailed
analysis of the effects of convolving the model predictions with
observational errors. All stellar masses were converted to those for a Salpeter
IMF.

We first compared the GSMFs derived for various redshifts in the upper panel of Fig. \ref{fig:mf_mod}; we
 plotted  only four redshift bins, the other two having very similar
behaviours to the 1.0 - 1.4 interval. We considered both
the BC03- and the CB07-based observed GSMFs. Despite the different
physical recipes adopted by the different semi-analytical models, their
predictions are remarkably similar \citep[as already found by
][]{fontanot09}. All the models considered consistently predict a
larger abundance of low mass galaxies than observations at
least up to $z\sim3$, despite the steep low-mass end slope inferred
from our data. The only exception is for the model of \cite{wang08}, which is closer than all other models to the observations at $z\sim 2.2$ and consistent with them at $z\sim 3$. 
The general overestimation of the faint-end is often attributed to a too efficient formation
of low-to-intermediate mass ($10^9-10^{11}M_\odot$) galaxies in the
models (\citealt{fontanot09}, see also \citealt{lofaro09}) and it cannot be explained by systematic uncertainties caused by the stellar
templates. On the observational side, it is unlikely that so many low
mass galaxies have been missed by observational surveys, especially at low redshift
($z\sim$ 1), when the disagreement is evident for $M_* >
10^{10}M_\odot$. In our highest redshift bin ($3.5<z<4.5$), the models of
\cite{wang08}, \cite{somerville11} and MORGANA show a reasonable
agreement with the data, while the \citet{menci06} model still
slightly overpredicts the space density of low mass galaxies. 
This is because the model of \cite{menci06}  includes starburst events triggered by fly-by interactions, which are very common in low mass objects at high redshift and increase their  stellar mass.  
All theoretical predictions also 
underestimate the stellar mass of high mass galaxies in the highest redshift bin. 
However, this mass
range is highly affected by cosmic variance, and the small ($\leq\ 3$)
number of galaxies observed with our data in each of the highest mass bins
($M>10^{11}M_\odot$) prevents us from drawing any firm conclusions. We note 
that \cite{marchesini09} found a similar disagreement at high masses,
using observations on a sky area $\sim 10$ times larger than the area
sampled by this study. 

A different treatment for the TP-AGB phase in the stellar population
synthesis models (i.e. moving from BC03 to CB07) may indeed help in reducing
the high-mass end discrepancy, by reducing the inferred stellar masses
of the galaxies in the sample.  \cite{henriques11} indeed demonstrated that the inclusion of a more detailed
treatment of the TP-AGB phase in the semi-analytical framework solves
a comparable mismatch between the predicted and observed rest-frame
$K$ band luminosity function of bright galaxies at $z\sim 2-3$, by
increasing the predicted $K$-band flux in model galaxies. This effect
is particularly relevant at these cosmic ages, since model galaxies
are dominated by intermediate-age stellar populations, with the right
age ($\sim 1$Gyr) to develop a TP-AGB phase, which makes them redder
without the need to change their mass or age. We test this idea by
considering stellar masses estimated using the CB07 synthetic templates.
However, as clearly shown in Fig. \ref{fig:mf_mod} (upper panel, red
open boxes), assuming stellar templates that account for TP-AGB stars
alleviates (the number of  $M>10^{11}M_\odot$ galaxies in the
sample is reduced by 37\%) but does not
completely solve the mismatch at the high-mass-end of the GSMF in the
highest redshift bin (where the previous percentage reduces to 12.5\%).

We also compared the various SMDs for galaxies of mass $10^9 < M_*/M_\odot < 10^{13}$, 
as predicted by the four models we consider (lower panel of
Fig. \ref{fig:mf_mod}), and we found that it is on average higher than
that observed by a factor of $\sim 2$ up to $z\sim 2$. 
The models of \cite{wang08}, \cite{somerville11}, and MORGANA 
show good agreement with our results at $2\lesssim z\lesssim 3$, and are below the data 
at the highest redshifts. 
The predictions of \cite{menci06}, given its overall steeper low-mass end, are
instead higher than the observed SMD up to $z\sim 3$. These results are driven by the afore mentioned overabundance of intermediate-to-low-mass galaxies, which dominate 
the SMD at all redshifts, and are counterbalanced by  the underestimation of the stellar mass in massive galaxies at high redshift.

\section{Summary}\label{sec:summ}

We have used deep near-IR observations ($Y$ as faint as 27.3, both $J$ and $H$ as faint as 27.4 AB mag at $5 \sigma$)  carried out with the Wide Field Camera 3 in the GOODS-S field as part of the Early Release Science. These data, complemented with a deep $K_S$ (as faint as 25.5 at $5 \sigma$) Hawk-I band data set and  high quality photometry in various bands from the near-UV to 8 \mic, have been used to derive accurate estimates of the stellar mass. We have succeeded in reducing the average relative error in the stellar  mass for a given object  by $\sim 30\%$ with respect to that obtained with the GOODS-MUSIC catalogue in previous works from our group.

Unfortunately, the sky region covered by ERS observations is biased by a number of localized as well as diffuse overdensities \citep{vanzella05,castellano07,salimbeni09,yang10,magliocchetti11}, which could be responsible for the oscillations in the normalization of the GSMF given the limited size of the area. 
However, this data set offers a unique combination of accuracy in photometric quality (hence in photometric redshifts and stellar mass) and depth, which makes it ideally suited to studying the faint-end of the GSMF.  

We computed the GSMFs in six different redshift intervals between 0.6 and 4.5. 
Thanks to the depth of the catalogue, we were able to study the low-mass end of the GSMF at lower masses than 
most previous studies by 0.5 dex up to $z\sim 1.8$ and 0.1 dex at $z\sim 4$. We compared our results with previous works and  found general good agreement, even at the highest masses, despite the limited sky area sampled by our data set. 
We found that the only redshift intervals that show poorer agreement with previous results are those between $z=1.4$ and $z=2.5$, which are known to be affected by the presence of overdensities: we discovered, as expected, a slightly larger abundance of massive galaxies.
We also compared our results  for the GSMF obtained with two different stellar libraries, BC03 templates and CB07 ones, the latter including a treatment of TP-AGB stars. 
The stellar masses inferred from the CB07 library are on average 0.12 dex lower than the BC03-based ones, with a large (0.17 dex) scatter. The lack of a clear trend with stellar mass or redshift of the ratio of the two estimates translates into a lack of  systematic difference between the best-fit Schechter parameters in the two cases. 
The largest disagreement was found at $z \sim 2 - 3$, where the effect of the TP-AGB phase is expected to be the most important. 

The main result of this study is the steepening of the faint-end slope: the value of $\alpha$ increases from $-1.44 \pm 0.03$ at $z\sim 0.8$ to $-1.86 \pm 0.16$ at $z\sim 3$, and then flattens up to $z\sim 4$. We have confirmed the steepening of the low-mass end, which had been pointed out by previous authors,  with deeper and higher quality photometry. 
Our results are unaffected by degeneracies in the $M^*$ parameter, and they are insensitive to the choice of either the stellar templates or the functional shape fitted to the GSMF, as well as to the limitations of the small area covered by ERS observations.

We computed the SMD as a function of redshift and compared it with the integrated star formation histories derived by \cite{hopkins06} and \cite{reddy09}. The finer  sampling of the GSMF at low masses and the steep inferred faint-end slopes  determine the higher SMD estimates at $z>2$ than most previous works, solving the disagreement observed by previous authors between the SMD and the integrated SFRD at these redshifts. However, 
despite the steep GSMF 
that we find, the integrated star formation history still exceeds the direct measure of the SMD at $z\sim 2$ by a factor of $\sim 2-3$, even when our data are analysed together with the results of previous large surveys to ensure a good sampling of also the bright-end tail of the GSMF. 

Finally, we compared our GSMF and SMD estimates with the predictions of four models of galaxy formation and evolution. All models predict a larger abundance of low mass galaxies than observations, at least up to $z\sim3$. They also underestimate the stellar mass of high mass galaxies 
in the highest redshift bin, although cosmic variance effects prevent us from drawing firm conclusions at these redshifts. 
The overabundance of low mass galaxies translates into a general overestimation of the total SMD with respect to the data 
 up to $z\sim 2$, while this density is underestimated at $z\gtrsim 3$  owing to the dearth of massive galaxies  at these redshifts. The exact degree of disagreement depends on the particular model.

Future CANDELS data will  cover a larger sky area and  allow a finer sampling of both the bright-end of the GSMF and  its normalization, and at the same time they will be deep enough to accurately probe the GSMF faint-end. These, together with spectroscopic follow-up campaigns, will reduce the uncertainties in the stellar masses, and they will significantly improve our results and our understanding of the stellar mass assembly process.

\begin{acknowledgements}
The authors thank M. Bolzonella, O. Ilbert, M. Kajisawa, and L. Pozzetti for providing their estimates of the stellar mass function, and N. Reddy for supplying the functional shape used to fit the SFRD.   
This work is based on observations made with the NASA/ESA Hubble Space Telescope,
obtained from the data archive at the Space Telescope Institute. STScI
is operated by the association of Universities for Research in
Astronomy, Inc. under the NASA contract NAS 5-26555.
Observations were also carried out using the Very Large Telescope at
the ESO Paranal Observatory under Programme IDs LP181.A-0717, 
ID 170.A-0788,  and the ESO Science Archive under Programme IDs 67.A-0249, 71.A-0584, and 
69.A-0539. 
We acknowledge financial contribution from the agreement ASI-INAF I/009/10/0".
\end{acknowledgements}


\end{document}